\documentclass[lettersize,journal]{IEEEtran}
\usepackage{amsmath,amsfonts}
\usepackage{algorithmic}
\usepackage{algorithm}
\usepackage{array}
\usepackage[caption=false,font=normalsize,labelfont=sf,textfont=sf]{subfig}
\usepackage{textcomp}
\usepackage{stfloats}
\usepackage{url}
\usepackage{verbatim}
\usepackage{graphicx}
\usepackage{ragged2e}
\usepackage{booktabs,enumitem}

\begin{document}

\title{Can Large Language Model Agents Balance Energy Systems?}

\author{%
    Xinxing Ren,  
    Chun Sing Lai,
    Gareth Taylor,  
    and Zekun Guo%
    \thanks{This work was supported in part by the IGNITE Network+, a UKRI research hub, grant ref EP/W033747/1 (Corresponding author: Zekun Guo).}%
    \thanks{Xinxing Ren, Chun Sing Lai, and Gareth Taylor are with the Department of Electronic and Electrical Engineering, Brunel University London, UB8\,3PH London, U.K. (e-mail: xinxing.ren@brunel.ac.uk; gareth.taylor@brunel.ac.uk).}%
    \thanks{Zekun Guo is with the Data Science, Artificial Intelligence and Modelling Centre, University of Hull, HU6 7RX Hull, U.K. (e-mail: z.guo2@hull.ac.uk).}%
}

\maketitle

\markboth{}%
{Shell \MakeLowercase{\textit{et al.}}: A Sample Article Using IEEEtran.cls for IEEE Journals}

\IEEEpubid{}

\maketitle

\begin{abstract}
This paper presents a hybrid approach that integrates Large Language Models (LLMs) with a multi-scenario Stochastic Unit Commitment (SUC) framework to enhance both efficiency and reliability under high wind generation uncertainties. In a 10-trial study on the test energy system, the traditional SUC approach incurs an average total cost of \$187.68 million, whereas the LLM-assisted SUC (LLM-SUC) achieves a mean cost of \$185.58 million (range: \$182.61--\$188.65 million), corresponding to a cost reduction of 1.1--2.7\%. Furthermore, LLM-SUC reduces load curtailment by 26.3\% (2.24 ± 0.31 GWh versus 3.04 GWh for SUC), while both methods maintain zero wind curtailment. Detailed temporal analysis shows that LLM-SUC achieves lower costs in the majority of time intervals and consistently outperforms SUC in 90\% of cases, with solutions clustering in a favorable cost-reliability region (Coefficient of Variation = 0.93\% for total cost and 13.8\% for load curtailment). By leveraging an LLM agent to guide generator commitment decisions and dynamically adjust to stochastic conditions, the proposed framework improves demand fulfillment and operational resilience.
\end{abstract}


\section{Introduction}
\IEEEPARstart{T}{he} increasing penetration of renewable energy (VRE) sources, such as wind and solar power, is revolutionizing modern power systems but also introducing significant operational challenges. Unlike conventional generation, which is dispatchable and predictable, VRE output is inherently uncertain and dependent on fluctuating weather conditions. This variability complicates traditional Unit Commitment (UC) practices, which were originally designed for more stable energy sources. To address these challenges, scenario-based methods have become a popular approach for handling uncertainty in VRE generation and system demand~\cite{Abujarad2017}. By generating multiple wind and load scenarios, these methods enable operators to capture stochastic variations and optimize generator commitments and dispatch decisions across a range of potential outcomes. Mixed-integer linear programming (MILP) is often employed for this purpose, offering a mathematically rigorous framework to solve such complex optimization problems.

While scenario-based MILP approaches have proven effective for UC problems, they suffer from significant scalability issues. As the number of scenarios grows to capture a wider range of uncertainties, the computational burden increases exponentially, driven by both the need to evaluate each scenario’s outcomes and the inherent combinatorial complexity of the binary commitment variables~\cite{Sturt2012}. To handle uncertainties in UC, chance-constrained MILP algorithms have been proposed, featuring convex formulations for Gaussian chance constraints and distributionally robust chance constraints~\cite{Badesa2023Chance}. However, finding optimal or near-optimal solutions within operationally tight timelines becomes increasingly challenging, especially in systems with high penetration of variable renewable energy (VRE).

In response to these challenges, researchers have turned to emerging technologies—most notably Artificial Intelligence (AI)—to complement traditional optimization techniques. Recent studies have investigated reinforcement learning (RL) models for solving UC problems~\cite{Omalley2023RL}. Although current results indicate that RL is decisively outperformed by well-established MILP methods, they also point to the potential for significant improvements that could make RL approaches more competitive. These developments illustrate how AI-driven methods may eventually enhance computational efficiency and improve decision-making under uncertainty in the UC domain.

In parallel, the development of Large Language Models (LLMs), such as OpenAI’s ChatGPT and GPT-4, has opened new possibilities for integrating advanced natural language processing (NLP) capabilities into technical domains. LLMs have demonstrated remarkable performance in tasks ranging from text generation and coding assistance to reasoning and context-aware problem-solving~\cite{Yao2023ReAct}. Beyond their conventional applications, there is growing interest in leveraging LLMs for tasks that require heuristic insights or domain-specific knowledge~\cite{Meng2024LLMA}. Some existing studies have explored LLM applications in power and energy systems. For example,~\cite{Huang2023} examines the use of foundation models like GPT-4 in power systems, highlighting their potential to optimize operational pipelines for tasks such as optimal power flow and scheduling problems. Additionally,~\cite{Li2024LLMImpacts} investigates the disruptive impacts of LLMs on power grids, focusing on their unique and transient energy consumption behaviours. In the context of power systems, LLMs’ ability to process vast amounts of contextual information and generate structured outputs offers a promising opportunity to complement traditional optimization techniques, particularly in areas like UC problem, where complexity and uncertainty pose significant challenges.

\begin{figure*}[!ht]
\centering
\begin{center}
\includegraphics[width=\textwidth]{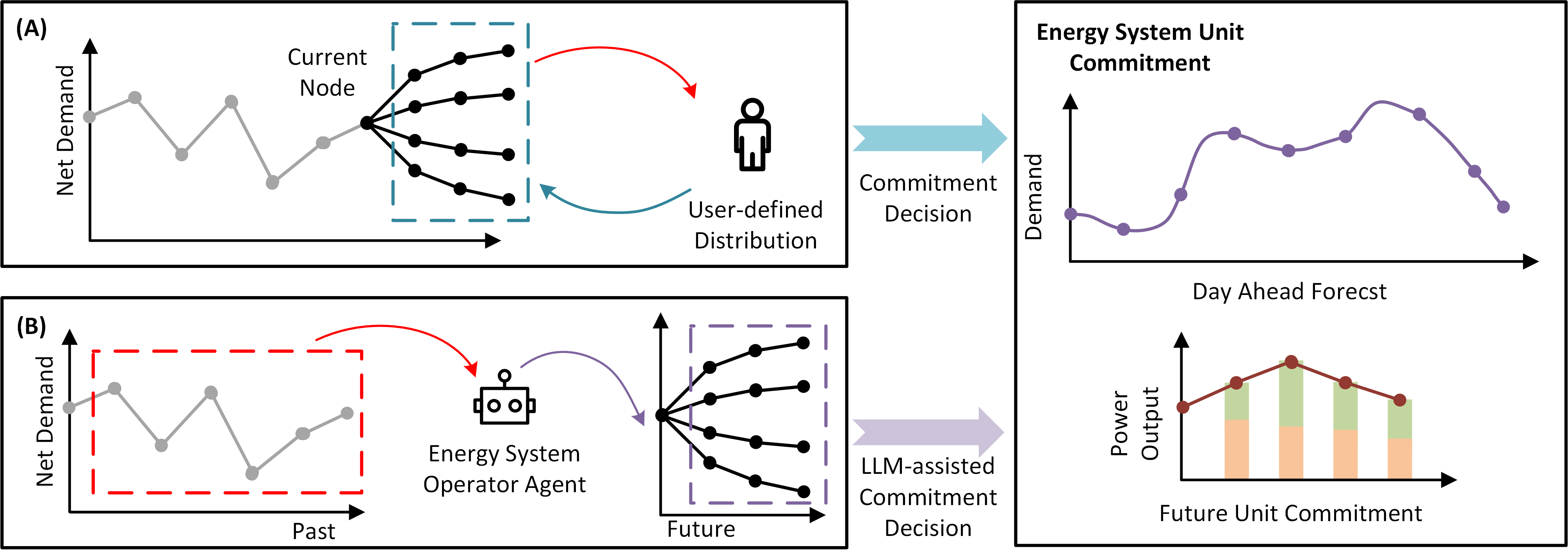}
\caption{Architecture diagram of LLM Agent-enabled energy system balancing framework. (A) User-defined balancing policy configuration interface. (B) Energy System Operator Agent decision process.}
\label{fig:framework}
\end{center}
\end{figure*}

In this paper, we investigate the novel application of an LLM to make decisions in a multi-scenario stochastic UC (SUC) problem. In this LLM-SUC framework, LLM agents can provide guidance to help the MILP to understand the uncertainties generated from wind energy. The proposed framework is a two-stage multi-agent system in which one agent interprets historical and predicted wind generation data, while the second agent derives the final output before feeding it back into the SUC algorithm.

\section{Methodology}
Recent advancements in LLMs have enabled applications beyond natural language processing, including optimization and scenario generation. Their capability to adaptively analyze and synthesize patterns provides a new avenue for balancing energy systems.

\subsection{Stochastic Unit Commitment}
The optimization objective for the multi-scenario, multi-stage model without energy storage is to minimize the expected total cost across all scenarios ($n$) and time steps ($t$), which includes startup costs, generation costs, load curtailment costs, and wind curtailment costs:
\begin{equation}
\min \sum_{t \in T} \sum_{n \in N} \pi(n) \Bigg(\sum_{g \in G} C_g(n,t) + \text{VOLL}\cdot P^{\text{Lcur}}(n,t)\Bigg)
\end{equation}
where $\pi(n)$ is the probability of scenario $n$, $C_g(n,t)$ is the operational cost of generating unit $g$ in scenario $n$ at time $t$, VOLL is the value of lost load, and $P^{\text{Lcur}}(n,t)$ is the amount of load curtailment in scenario $n$ at time $t$.

The operating cost of generating unit $g$:
\begin{equation}
C_g(n,t) = c_g^{\text{st}} N_g^{\text{sg}}(n,t) + \Delta \tau\Big(c_g^T P_g(n,t)\Big)
\end{equation}
where $c_g^{\text{st}}$ is the startup cost of generating unit $g$, $N_g^{\text{sg}}(n,t)$ is the startup decision for scenario $n$ at time $t$, $\Delta \tau$ is the time duration, $c_g^T$ is the generation cost per unit of power, and $P_g(n,t)$ is the generation output for scenario $n$ at time $t$.

\subsection{Problem Formulation}
The stochastic unit commitment problem is subject to several operational and physical constraints to ensure feasibility and reliability. Key constraints include:

\subsubsection{Power Balance Constraint}
\begin{equation}
\sum_{g=1}^{N_{\text{gen}}} P_g(n,t) + P^{\text{Wind}}(n,t) - P^{\text{Wcur}}(n,t) = P^{\text{L}}(n,t) - P^{\text{Lcur}}(n,t)
\end{equation}
where $P^{\text{Wind}}(n,t)$ is the wind power generation, $P^{\text{Wcur}}(n,t)$ is the wind curtailment, and $P^{\text{L}}(n,t)$ is the system demand at time $t$.

\subsubsection{Generation Limits}
\begin{equation}
 y_g(n,t) \cdot P_g^{\min} \leq P_g(t) \leq y_g(n,t) \cdot P_g^{\max}
\end{equation}

\subsubsection{Startup and Shutdown Constraints}
\begin{equation}
N_g^{\text{sg}}(n,t) \geq y_g(n,t) - y_g(n,t-1)
\end{equation}

\subsubsection{Ramping Constraints}
\begin{equation}
\Delta \tau\cdot RD\cdot y_g(n,t-1) \leq P_g(t) - P_g(t-1) \leq \Delta \tau\cdot RU\cdot y_g(n,t-1)
\end{equation}
where $RU$ and $RD$ are the upward and downward ramping limits of generator $g$.

\subsubsection{Curtailment Limits}
\begin{equation}
0 \leq P^{\text{Wcur}}(n,t) \leq P^{\text{Wind}}(n,t)
\end{equation}
\begin{equation}
0 \leq P^{\text{Lcur}}(n,t) \leq P^{\text{L}}(n,t)
\end{equation}

\subsection{LLM Agent and Scenario Tree Generation}
The proposed model leverages a Large Language Model (LLM) agent to facilitate scenario tree generation for addressing uncertainties in wind power forecasts. Scenario trees are essential for modeling variations in wind generation over multiple time steps, enabling robust decision-making in high-renewable energy systems. The LLM agent enhances the process by dynamically adjusting quantile parameters, interpreting wind error distributions, and automating key computational tasks.

\begin{figure}[!h]
\centering
\includegraphics[width=\linewidth]{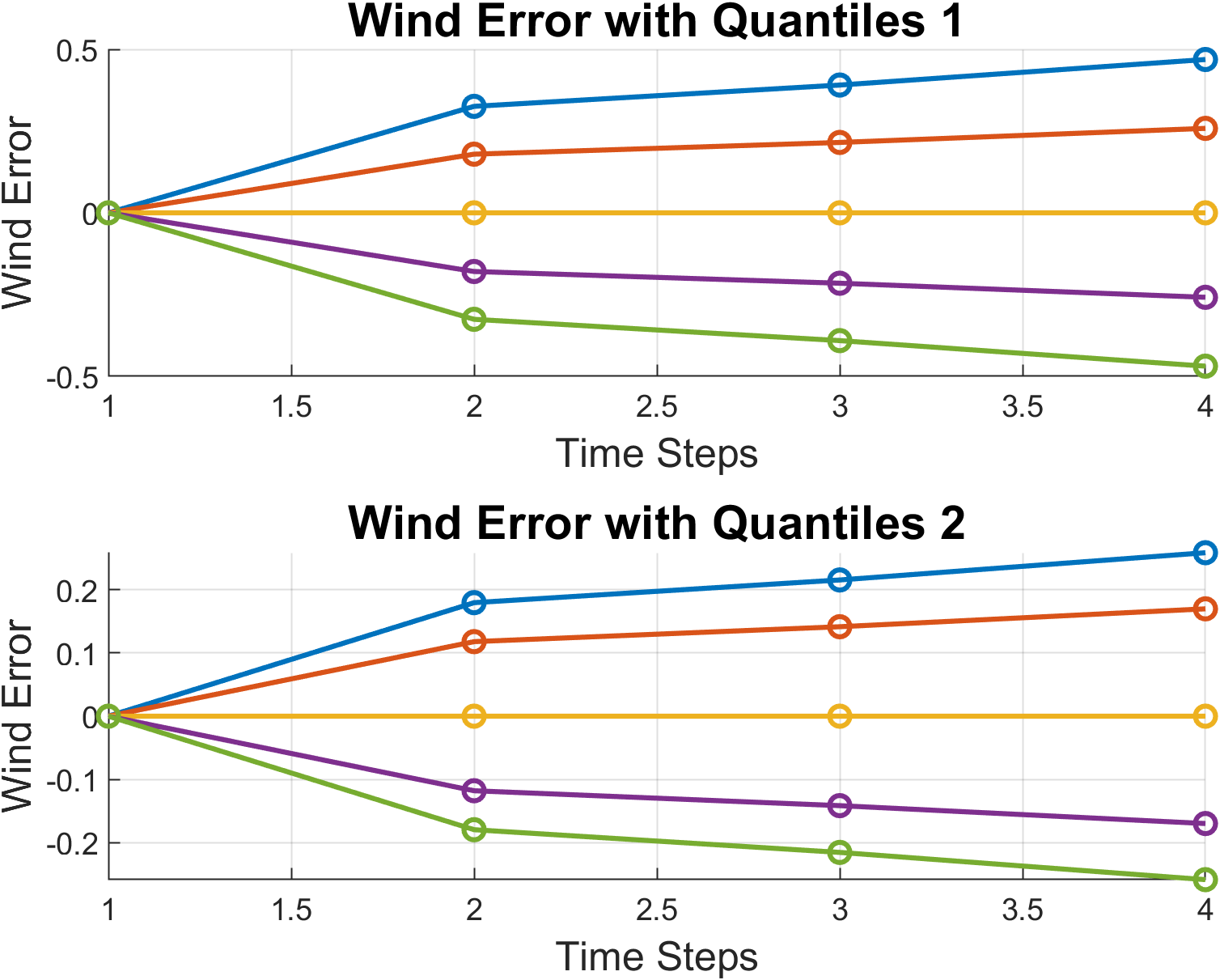}
\caption{Wind error under different quantile settings.}
\label{fig:quantiles}
\end{figure}

\textbf{Quantile Adjustment:} 
At each step, the LLM agent analyzes current wind forecast and actual values, past predictions and realisations, and future forecast scenarios. Figure~\ref{fig:quantiles} shows wind errors under different quantile settings.

In this work, wind forecast errors are modeled using an autoregressive (AR(1)) process with parameters for persistence ($\phi = 1.2$) and error scaling ($\epsilon_c = 0.14$), this scenario tree generation method was developed by ~\cite{Sturt2012} and ~\cite{Badesa2019}. Quantiles, for example $[0.99, 0.9, 0.5, 0.1, 0.01]$, define the scenarios at each node of the tree. The wind error at each time step is recursively calculated as:
\begin{equation}
W_e(n,k) = 
\begin{cases}
0,& k=1,\\
\phi \cdot W_e(n,k-1) + \epsilon_c \Phi^{-1}(\text{quantiles}), & k=2,\\
\phi \cdot W_e(n,k-1), & k>2,
\end{cases}
\end{equation}
where $W_e(n,k)$ is the wind forecast error for scenario $n$ at time $t$, $\epsilon_c$ is the wind forecast error scaling factor, $\phi$ is the parameter for persistence, and $\Phi^{-1}$ is the inverse cumulative distribution function.

Probabilities for reaching each node are calculated based on quantile intervals, ensuring accurate representation of wind error distributions. Figure~\ref{fig:quantiles} illustrates the wind error dynamics for two quantiles (Quantile 1 [0.01, 0.1, 0.5, 0.9, 0.99] and Quantile 2 [0.1, 0.2, 0.5, 0.8, 0.9]), highlighting their divergence over time.

\subsubsection{Quantile-Based Scenarios}
\label{sec:llm-prompts}
Quantiles $q \in [0.01, 0.1, 0.5, 0.9, 0.99]$ define scenario branches. Wind forecast errors $e(n,t)$ follow:
\begin{equation}
e(n,t) = \varphi \cdot e(n,t-1) + \epsilon_C \cdot \text{norminv}(q, \mu, \sigma)
\end{equation}
where $\varphi$ is the autoregressive parameter, and $\epsilon_C$ scales the uncertainty.

\subsubsection{LLM-Driven Refinement via Prompting}
\label{sec:llm-prompts}
This scenario tree generation, combined with the LLM agent’s interpretative capabilities, creates a powerful tool for enhancing energy system awareness and optimizing renewable integration. The LLM agents utilized in this study were developed based on the Microsoft Autogen framework, with two LLM agents defined to ensure accurate and desired outputs. Together, these prompts guide the LLM agents through a multi-step pipeline:

\noindent

\begin{itemize}
\item{\textbf{Agent 1:} Identifies pertinent quantiles by analysing forecast error statistics.}
\item{\textbf{Agent 2:}  Constructs a refined probability vector that may be asymmetric or otherwise altered, ensuring that scenario trees reflect real-world forecasting biases.}
\end{itemize}

The first LLM agent was guided by a structured prompt specifically designed to address uncertainties in power system modeling, particularly related to wind power and load forecasting errors. The task assigned to the LLM was as follows:

\begin{table}[!h]
\centering
\label{tab:agent1}

{\fontsize{10pt}{12pt}\selectfont
\begin{tabular}{@{}p{0.98\linewidth}@{}}
\toprule
{\raggedright
\textbf{Task Description for Agent 1:}
}\\[4pt]
\midrule
{\justifying
\noindent
The task involves constructing a wind power scenario tree based on an AR(1) process. You are given two sets of data: one representing the actual power demand and the other representing forecasted power demand. The error between these values (i.e., actual value - forecasted value) follows a normal distribution with a mean of 0 and a standard deviation of 1. We have five fixed quantiles [0.01, 0.1, 0.5, 0.9, 0.99] representing different possible deviations (errors) in wind power forecasts. Initially, the code assigns a probability vector of [0.05556, 0.24444, 0.4, 0.24444, 0.05556] to these five branches.

\noindent
Your task is to revise these probabilities based on historical forecast error data. The new probability distribution \texttt{prob\_new} must still sum to 1 but can be shifted to better reflect observed biases (e.g., if real wind power tends to exceed or fall below the forecast more often than expected). The output should include:

\begin{itemize}
\item{A new probability vector \texttt{prob\_new} associated with the same five quantiles remains non-negative and sums to 1.}
\item{A justification or reasoning for how that new distribution was derived (which may involve statistical calibration or heuristic adjustments).}
\item{An explanation of why the revised probabilities may yield more realistic or robust outcomes for wind power planning compared to the original symmetrical distribution.}
\end{itemize}

}\\
\bottomrule
\end{tabular}
}
\end{table}

\subsubsection{Example Provided in the Prompt}
\label{sec:prompt-example}

\noindent
The prompt included a sample query showcasing error calculations, 
mean and standard deviation derivations, and a step-by-step quantile selection process. 
For example:

\begin{enumerate}[label=\alph*),leftmargin=*]
\item \textbf{Compute the errors:} For each pair of actual and forecasted values, calculate the error:
\begin{equation}
e(t) = \text{Actual}(t) - \text{Forecasted}(t)
\end{equation}

\item \textbf{Derive the mean and standard deviation of the errors:}
Use the formulas:
\begin{equation}
\mu = \frac{\sum e(t)}{N}
\end{equation}
\begin{equation}
\sigma = \sqrt{\frac{\sum (e(t) - \mu)^2}{N - 1}}
\end{equation}

\end{enumerate}

Based on the output of the first agent, another agent was defined to generate a list of new probabilities. The following prompt was used to generate the list of probabilities.

\begin{table}[!h]
\centering
\label{tab:agent1}

{\fontsize{10pt}{12pt}\selectfont
\begin{tabular}{@{}p{0.98\linewidth}@{}}
\toprule
{\raggedright
\textbf{Task Description for Agent 2:}
}\\[4pt]
\midrule
{\justifying
\noindent
Your task is to extract 'the prob\_new finally selected' from the given text. Search for the exact phrase to locate the probability value. The final output should strictly follow this JSON format:
\{ 'prob\_new': [p1, p2, p3, p4, p5] \}, where value is the extracted probability. For example:

\noindent
\textbf{Input:} 
'The calculation resulted in the prob\_new finally selected: 0.05, 0.25, 0.4, 0.25, 0.05.'

\noindent
\textbf{Output:} 
\{ 'prob\_new': [0.05, 0.25, 0.4, 0.25, 0.05] \}

\noindent
Make sure to extract only the specified value and return it in the correct format.
}\\
\bottomrule
\end{tabular}
}
\end{table}

This two-agent framework, based on the Microsoft Autogen environment, ensures that each agent performs a focused task: the first handles statistical computations and adjusts probability distributions based on those computations, the second ensures the output aligns with the probability requirements. This multi-agent pipeline enhances the transparency and robustness of the output, enabling researchers to inspect each stage of the prompt-driven process and validate the resulting probability vectors against real-world performance criteria. This approach leverages the LLM's reasoning capabilities for tasks that require domain-specific interpretations and precise numerical outputs.

Finally, the refined new probabilities are used as input to the SUC algorithm to compute predictive commitment results. For each time step, the LLM-SUC executes its calculations iteratively.

\section{Case Studies}
The proposed work addresses the challenge of balancing a high-renewable energy system by leveraging a Large Language Model (LLM) agent to enhance system awareness and decision-making. This innovative approach integrates advanced LLM capabilities with power system modeling, enabling dynamic and informed responses to uncertainties such as renewable energy fluctuations and demand variability.

To demonstrate its effectiveness, two case studies were conducted on high-renewable energy systems. These case studies highlight the ability of the LLM-SUC model to optimize system operations, improve cost-efficiency, and ensure reliability. The optimization problem, formulated as a mixed-integer linear program (MILP), was solved using Gurobi and YALMIP within the MATLAB environment, showcasing robust computational performance.

The LLM used in this work was OpenAI's GPT-4, which played a pivotal role in enhancing the model's interpretability and adaptability by analyzing scenarios, refining optimization parameters, and guiding decision-making. The results validate the proposed model's capability to balance renewable-dominated power systems effectively, paving the way for integrating LLM agentic methodologies in energy system operations.

\subsection{Energy System Setup}

This study investigates an energy system composed of three conventional generators, wind power generation, and load demand, optimized under a rolling horizon framework. The load demand data, with a peak of 38 GW and a minimum of 18 GW, is derived from historical records and scaled to represent realistic consumption. Wind power is modeled using real-time data for the first hour and forecasted values for the next four hours, with uncertainties captured using a scenario tree. Wind generation is capped at 20 GW to reflect physical constraints. The thermal generators have defined capacity limits, startup costs, and ramping constraints, ensuring operational realism. Load curtailment and wind curtailment mechanisms are incorporated to handle supply-demand imbalances. Load curtailment is penalized with a high value of lost load (VOLL) to reflect its economic impact, while wind curtailment addresses excess generation beyond system capacity. The VOLL was set as 300,000 \$/GWh. The other settings for the energy system are shown in Table~\ref{table:gen_settings}.

\begin{table}[!h]
\centering
\caption{Energy Generator Settings for Case Studies}
\begin{tabular}{@{}lccc@{}}
\toprule
\textbf{Parameters} & \textbf{Gen Unit 1} & \textbf{Gen Unit 2} & \textbf{Gen Unit 3} \\
\midrule
Startup cost (million \$) & 4 & 2 & 4 \\
Max power (GW) & 10 & 12 & 15 \\
Min generation (GW) & 3 & 2 & 0 \\
Generation cost (1000\$/GWh) & 40 & 60 & 120 \\
Ramp up/down limit (GW/h) & 4 & 4 & 6 \\
\bottomrule
\end{tabular}
\label{table:gen_settings}
\end{table}

A rolling optimization framework with a 4-hour forecast horizon is employed, solving for each hour over a 24-hour simulation period. Mixed-integer linear programming (MILP) is used to minimize costs, including generation, startup, and curtailment. This setup ensures reliable operation under uncertain wind conditions, with results capturing generator outputs, wind utilization, and curtailment decisions. The model highlights the trade-offs in balancing cost and system reliability without energy storage.

\subsection{LLM Model Selection Rationale}
GPT-4 was selected as the core LLM model due to its:
\begin{itemize}
    \item Superior accuracy in statistical reasoning tasks compared to GPT-3.5 or other open-weights models.
    \item Improved performance in preliminary experiments, demonstrating precise probability adjustment capabilities critical for scenario-tree generation.
    \item Robust multi-step reasoning suitable for complex energy decision-making scenarios.
\end{itemize}

Future research could further evaluate and compare other models such as GPT-4.5, GPT-4-turbo, Google Gemini, or Anthropic Claude, considering their potential trade-offs in accuracy, speed, and cost.

\subsection{Stochastic Unit Commitment Results}
\begin{figure}[!h]
\centering
\includegraphics[width=\linewidth]{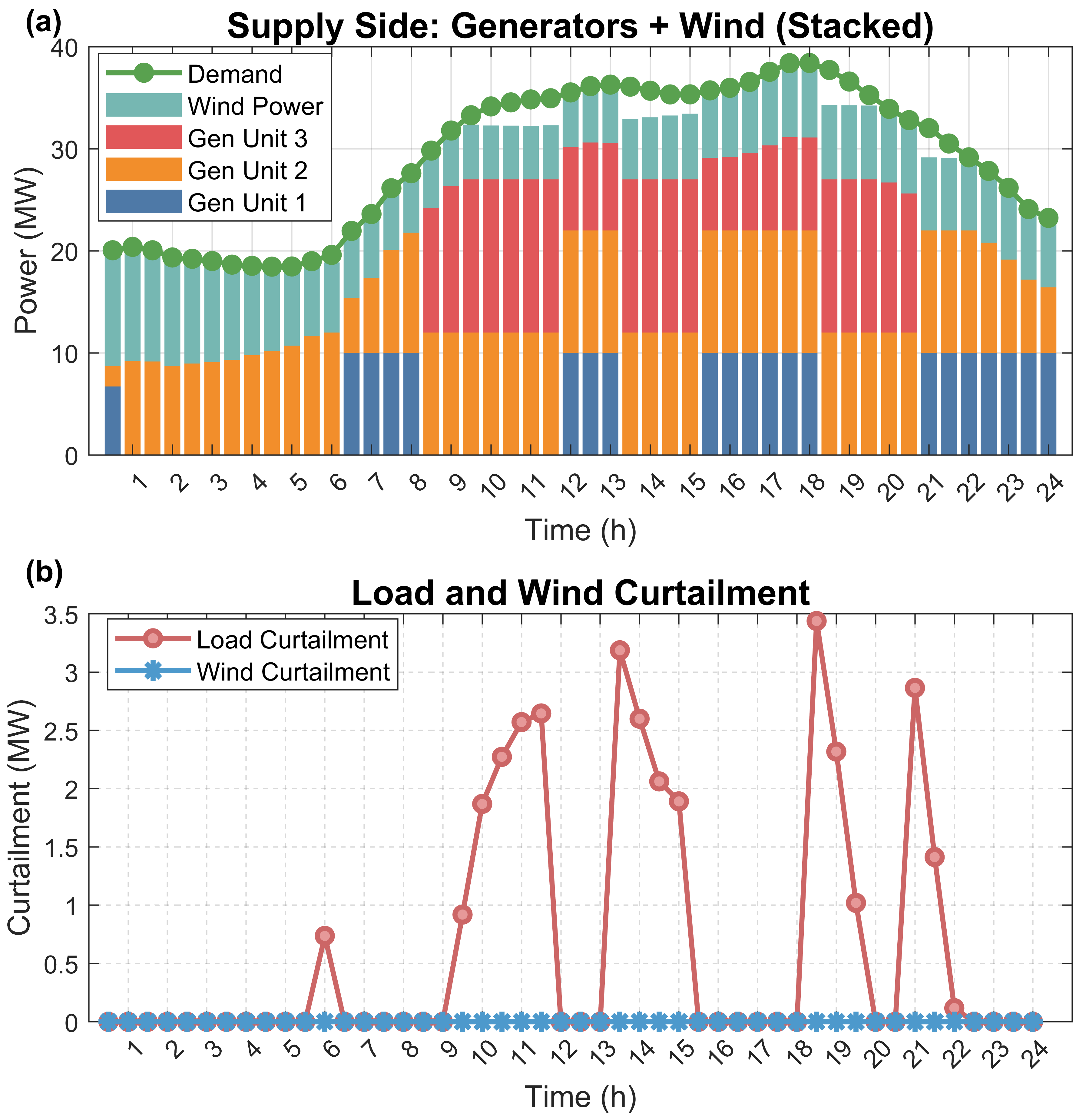}
\caption{Energy dispatch and load curtailment by stochastic unit commitment. (a) Generation dispatch. (b) Load and wind curtailment.}
\label{fig:suc_results}
\end{figure}

Fig.~\ref{fig:suc_results} illustrates the results of stochastic unit commitment (SUC) optimization for a typical day, showcasing energy dispatch and curtailment strategies under uncertain conditions. Fig.~\ref{fig:suc_results}(a) presents the stacked generation output of multiple units and wind power compared against the total system demand over a 24-hour horizon. The figure highlights the dynamic interplay between conventional generation (Gen Unit 1–3) and renewable wind power, with wind output peaking early in the period. The alignment of supply with demand, indicated by the total stacked output meeting the demand curve, demonstrates effective dispatch planning. Fig.~\ref{fig:suc_results}(b) shows the load and wind curtailment across the same period, with significant curtailment events occurring during periods of high wind power output.

\begin{figure}[!h]
\centering
\includegraphics[width=\linewidth]{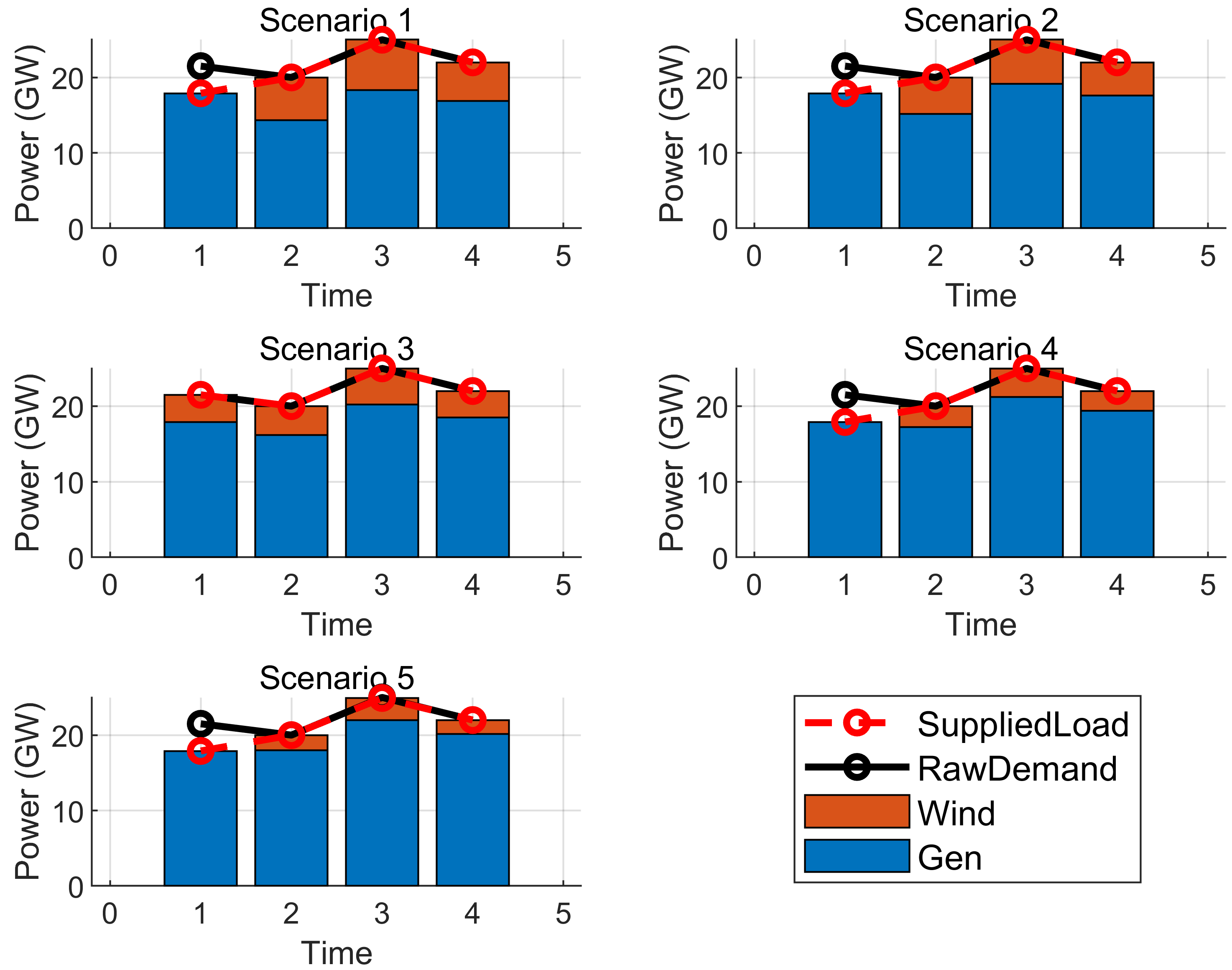}
\caption{Energy dispatch decision-making for different scenarios.}
\label{fig:multi_scenario}
\end{figure}

\begin{figure}[!h]
\centering
\includegraphics[width=\linewidth]{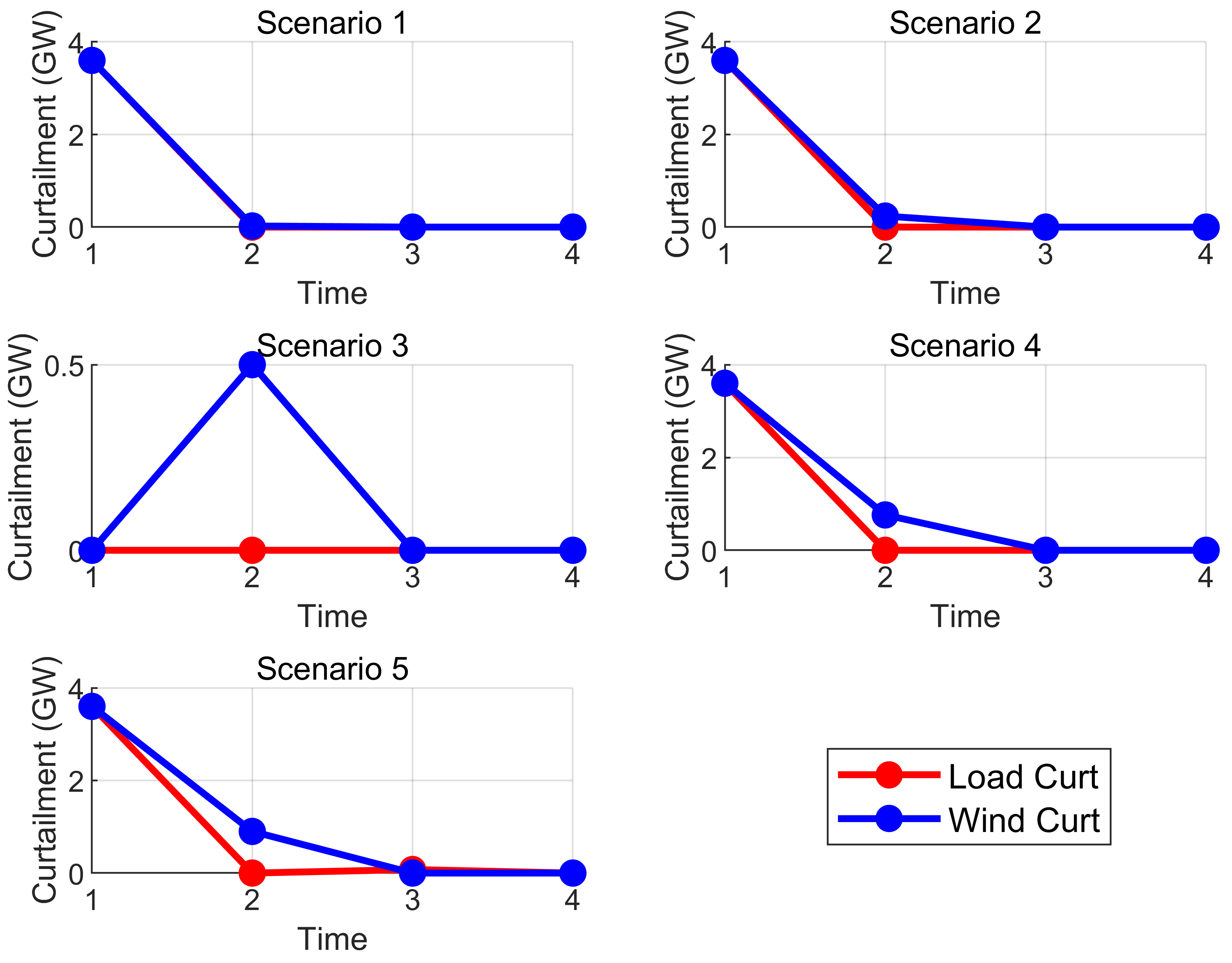}
\caption{Curtailment decision-making for different scenarios.}
\label{fig:curtailment_scenarios}
\end{figure}

Fig.~\ref{fig:multi_scenario} illustrates the energy dispatch decisions across five scenarios, showcasing how conventional generation and wind power are dynamically adjusted to meet demand at each time step. The stacked bar charts represent the contribution of generation units and wind power, while the line plots compare raw demand with the actual supplied load after accounting for load curtailments. For example, in Scenario 1 at time step 1, there is a clear mismatch between raw demand and supplied load due to limited generation and reduced wind power availability, highlighting the need for load curtailment.
Fig.~\ref{fig:curtailment_scenarios} illustrates the curtailment decisions for load and wind power across various scenarios. In this figure, the red line represents load curtailment while the blue line shows wind curtailment. For example, in Scenario 1 at time step 1, there is a noticeable amount of wind curtailment—this reflects a situation where the available wind power exceeds the system’s ability to absorb it. On the other hand, Scenario 4 at time step 1 shows that both load and wind curtailments occur simultaneously, which indicates a period of operational difficulty where both supply and demand sides require adjustment. These results demonstrate that the SUC framework is capable of managing uncertainties effectively by adjusting curtailment strategies to balance supply and demand.

\subsection{Economic and Performance Analysis}
\begin{figure}[!h]
\centering
\includegraphics[width=\linewidth]{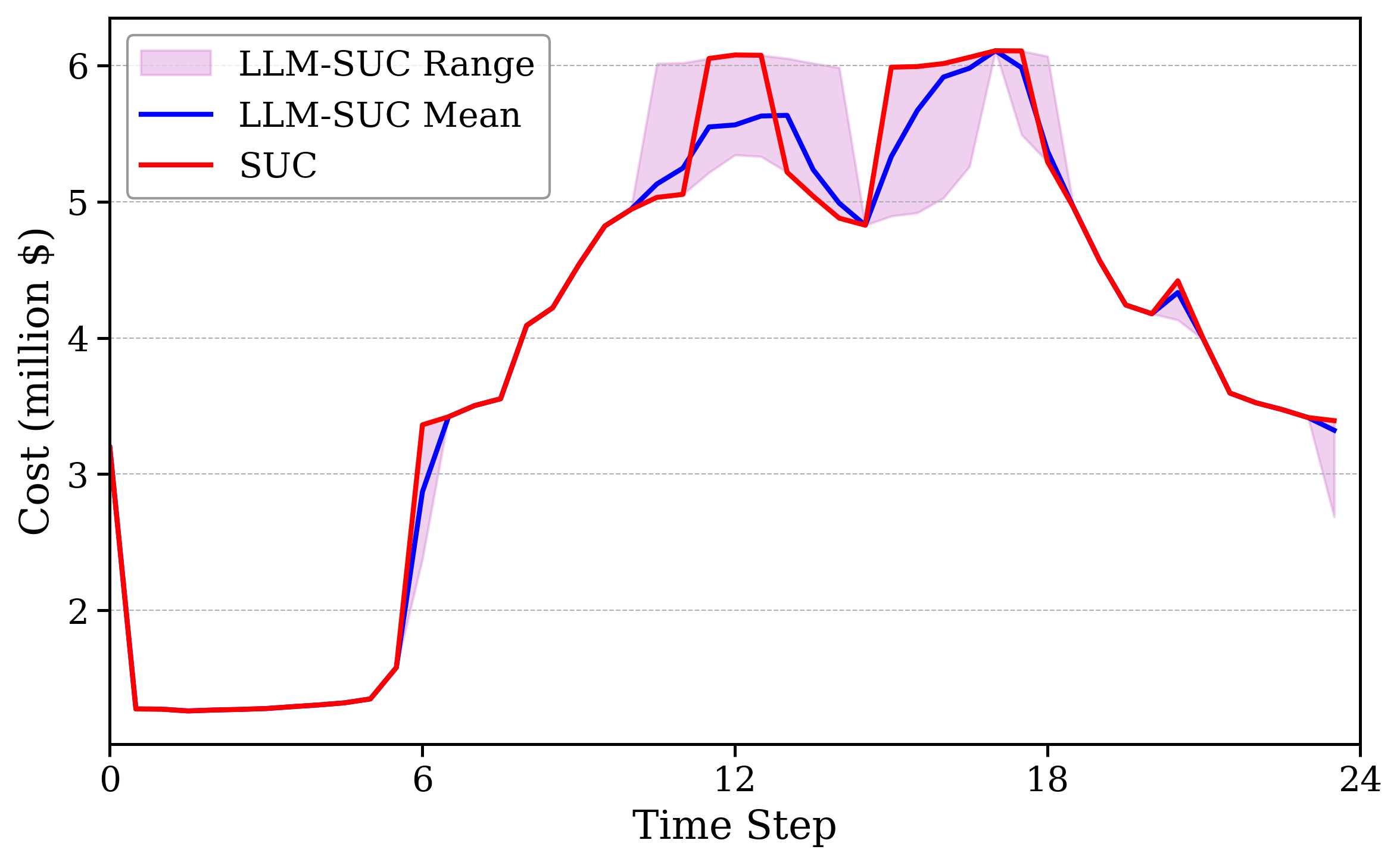}
\captionsetup{skip=0pt}
\caption{Daily cost comparison across 48 half-hour intervals, showing SUC versus LLM-SUC mean and range.}
\label{fig:day_stochastic}
\end{figure}

\begin{figure}[!h]
\centering
\includegraphics[width=\linewidth]{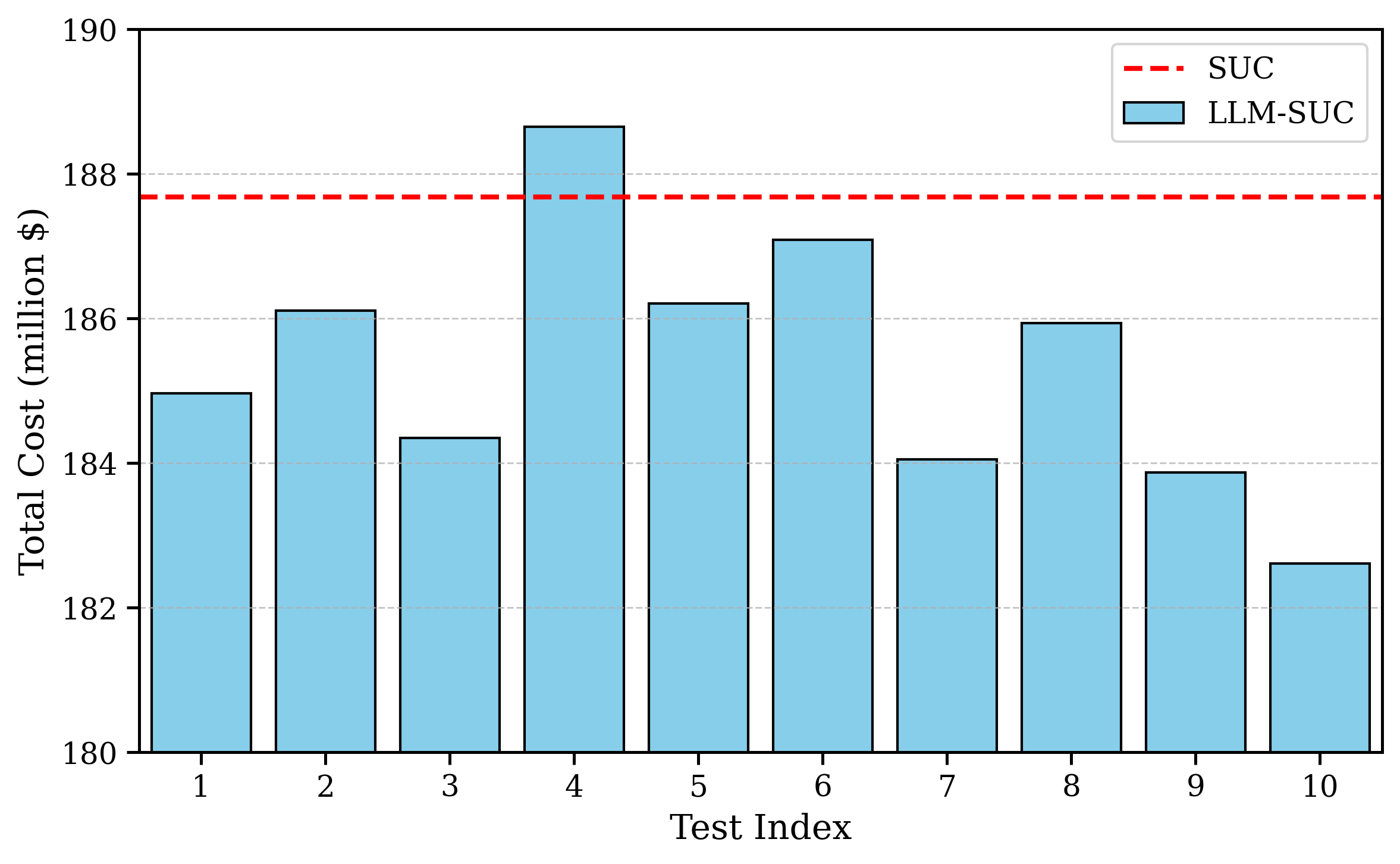}
\captionsetup{skip=0pt}
\caption{Distribution of daily total costs from 10 LLM-SUC trials compared to SUC baseline (red line).}
\label{fig:cost_compare}
\end{figure}

Fig.~\ref{fig:day_stochastic} and Fig.~\ref{fig:cost_compare} provide a detailed comparison between 10 independent LLM-SUC trials and the traditional SUC approach. The LLM-SUC model achieves a mean total daily cost of \$185.58 million (with a range spanning from \$182.61 million to \$188.65 million), representing a cost reduction between 1.1\% and 2.7\% compared to the SUC baseline of \$187.68 million. Table~\ref{table:results_uc} further highlights that LLM-SUC reduces load curtailment by 26.3\% (2.24 ± 0.31~GWh versus 3.04~GWh for SUC) while maintaining zero wind curtailment.

\begin{table}[!h]
\centering
\caption{Comparative Performance Results}
\label{table:results_uc}
{\small
\begin{tabular}{l c c}
\toprule
\textbf{Metric} & \textbf{SUC} & \textbf{LLM-SUC (10 trials)} \\
\midrule
Total Cost (M\$) & 187.68 & 185.58 ± 1.72 \\
Load Curtailment (GWh) & 3.04 & 2.24 ± 0.31 \\
Wind Curtailment (GWh) & 0 & 0 \\
\bottomrule
\end{tabular}
}
\end{table}

Fig.~\ref{fig:day_stochastic} shows the temporal performance over 48 rolling steps (corresponding to 24 hours). In the LLM-SUC results, the cost range (represented by the filled region) varies considerably over time—from as low as approximately 1.26 in the early hours to over 6.11 in the peak periods—indicating that the stochastic adjustments allow the model to exploit favorable conditions at different times. In contrast, the traditional SUC cost profile is relatively flat, suggesting that it may not capture or adapt to temporal variations in system conditions as effectively.

Fig.~\ref{fig:cost_compare} compares the overall daily cost distributions. In 9 out of 10 trials, the LLM-SUC trials yield lower daily costs than the SUC baseline (red line at \$187.68 million). The LLM-SUC cost values, which range from \$182.61 million to \$188.65 million, indicate that while one trial produced a higher cost, the majority of solutions cluster in a more favorable cost region. This clustering, along with the cost reduction, underscores LLM-SUC's ability to better manage uncertainties and optimize the system's operation.

\subsection{Analysis of LLM Output Stability}
The 10-trial experiment reveals moderate cost variability ($\sigma$ = 1.72M\$; Coefficient of Variation = 0.93\%) and a 90\% success rate relative to the SUC baseline. Load curtailment is notably consistent ($\sigma$ = 0.31GWh; Coefficient of Variation = 13.8\%), while wind curtailment remains at zero across all trials. Although the model sometimes produces simultaneous load and wind curtailment in isolated instances, this phenomenon can be explained by the lack of an explicit constraint to preclude simultaneous curtailment. Instead, the model minimizes the overall cost by balancing the high penalty on load curtailment (through VOLL) with the absence of a cost for wind curtailment. The resulting controlled randomness in the LLM-generated scenario tree leads to slight variations in the probability vectors between trials while preserving a coherent adjustment pattern that sustains solution quality.

\section{Conclusions}
This study demonstrates that the LLM-SUC model can reduce total daily costs by 1.1–2.7\% and lower load curtailment by 26.3\% compared to traditional SUC, while maintaining zero wind curtailment. The LLM-SUC results exhibit a more dynamic cost profile, with the cost range varying significantly throughout the day, thereby capturing the temporal variability in system conditions. Although the model sometimes exhibits simultaneous load and wind curtailment, this is attributable to the absence of a constraint explicitly forbidding such a combination. Instead, the model balances supply and demand through its cost function, which heavily penalizes load curtailment while leaving wind curtailment unconstrained. The 10-trial analysis shows that LLM-SUC consistently outperforms traditional SUC in 90\% of cases, with solutions clustering in a favorable cost-reliability Pareto region. The median LLM-SUC solution translates to approximately \$2.1 million in daily savings. Future work should explore enhanced prompt engineering and additional constraints to further refine the curtailment decisions, thereby reducing the incidence of simultaneous curtailments while maintaining robust performance improvements.

\bibliographystyle{IEEEtran}
\bibliography{ref.bib}

\newpage
 
\vspace{11pt}

\vfill

\end{document}